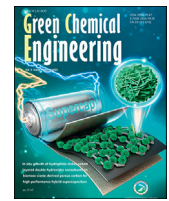

Article

# In situ growth of hydrophilic nickel-cobalt layered double hydroxides nanosheets on biomass waste-derived porous carbon for high-performance hybrid supercapacitors

Yuchen Wang [a], Yaoyu Liu [a], Zuo Chen [a], Man Zhang [a], Biying Liu [a], Zhenhao Xu [a], Kai Yan [a,b,*]

[a] School of Environmental Science and Engineering, Sun Yat-sen University, 135 Xingang Xi Road, Guangzhou, 510275, China
[b] Guangdong Provincial Key Laboratory of Environmental Pollution Control and Remediation Technology, Guangzhou, 510275, China

## HIGHLIGHTS

- NiCo-LDHs nanosheets on biomass waste-derived porous carbon (BC) were achieved using *in situ* growth method.
- *In situ* growth process under ultra-sonication realizes the rational arrangement of NiCo-LDHs nanosheets.
- NiCo-LDHs/BC material possesses an ultra-high specific capacitance of 2390 F g$^{-1}$ (956 C g$^{-1}$) at 1 A g$^{-1}$.
- The high electrochemical performance is attributed to large surface area, high conductivity, and enhanced wettability.

## GRAPHICAL ABSTRACT

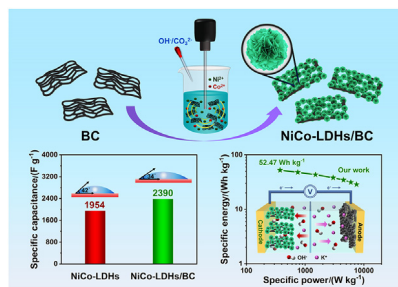

## ARTICLE INFO

## ABSTRACT

Rational design and cost-effective fabrication of layered double hydroxides (LDHs) nanosheets with extraordinary electrochemical performance is a key challenge for hybrid supercapacitors (HSCs). Herein, we report a facile *in situ* growth methodology to eco-friendly synthesize hydrophilic NiCo-LDHs nanosheets on biomass waste-derived porous carbon (BC) for robust high-performance HSC cathode. The *in situ* growth process under ultrasonication realizes the rational arrangement of NiCo-LDHs nanosheets on the surface of BC, which effectively increases the specific surface area, promotes the electronic conductivity and enhances the wettability of NiCo-LDHs nanosheets without affecting their thickness values. With the beneficial effects of ultrathin thickness of LDHs nanosheets (6.20 nm), large specific surface area (2324.1 m$^2$ g$^{-1}$), low charge transfer resistance (1.65 Ω), and high wettability with electrolyte (34°–35°), the obtained Ni$_2$Co$_1$-LDHs/BC50 electrode possesses an ultra-high specific capacitance of 2390 F g$^{-1}$ (956 C g$^{-1}$) at 1 A g$^{-1}$, which is superior to most reported values. Furthermore, an assembled Ni$_2$Co$_1$-LDHs/BC50//YP-80F HSC delivers a maximum specific energy of 52.47 Wh kg$^{-1}$ at 375 W kg$^{-1}$, and maintains a high capacitance retention of 75.9% even after 4000 cycles. This work provides a facile approach to fabricate LDHs nanosheets based cathode materials for high-performance HSCs.

## 1. Introduction

The depletion of fossil fuels and rapid climate changes accelerate the development of high-performance energy storage devices [1,2]. Supercapacitors (SCs) have attracted tremendous attention due to their fast charge/discharge capability, high specific power, and prominent cycle stability [3,4]. However, the intrinsically low specific energy hinders the widespread utilization of SCs as stand-alone applications [5,6]. To overcome this barrier, a hybrid supercapacitor (HSC) with a Faradaic electrode and a capacitive electrode has been proposed to achieve high specific power and high specific energy simultaneously [7–9]. Hence, the selection of electrode materials is of vital importance in deciding electrochemical performance of HSCs.

Layered double hydroxides (LDHs) are considered as excellent cathode materials for HSCs with advantages, including large surface area and controllable layered structure [10,11]. Nevertheless, LDHs, especially bulky LDHs, are suffered from low electronic conductivity and serious aggregation, leading to undesirable electrochemical performance [12,13]. Up to date, the mostly common strategy is to delaminate LDHs into mono-layered or few-layered LDHs nanosheets through liquid exfoliation [14,15], ultrasonication [16,17], etching [18,19], etc. These exfoliation methods could enhance the exposed surface area and elevate electronic conductivity of LDHs materials, which are beneficial for boosting energy storage.

Recent research works have utilized carbon materials including porous carbon (PC), graphene, reduced graphene oxide (rGO), and carbon nanotube to further promote the rate capability and charge capacity of LDHs nanosheets [20,21]. For example, Yu et al. [22] grew NiMn-LDHs nanosheets on zeolitic imidazolate frameworks-8 metal-organic frameworks (ZIF-8 MOFs) derived PC to reduce the agglomeration and enable fast ion transport, leading to the excellent specific capacitance of 1634 F g$^{-1}$ at 1 A g$^{-1}$, which was almost 1.5 times higher than that of pristine NiMn-LDHs nanosheets. Successively, Han and Dai et al. [23] fabricated NiCoAl-LDHs nanosheets on rGO *via* a facile static assembly strategy. The electronic coupling interaction between LDHs nanosheets and rGO displays a beneficial synergetic effect on enhancing the capacitive performance of LDHs nanosheets. The resulting composites possessed a high specific capacitance of 1877 F g$^{-1}$ at 1 A g$^{-1}$, which was much greater than that of pristine NiCoAl-LDHs nanosheets (1367 F g$^{-1}$). Despite the prominent improvements of these carbon materials, their current utilization still faces the dilemma of the relatively high manufacturing costs and toxic organic solvents. Additionally, the boundary lubrication between LDHs nanosheets and these carbon materials is not desirable. Recently, biomass derived-carbon materials have drawn tremendous interests as electrode materials of SCs due to low cost, environmental friendliness, hierarchical porous structure and high hydrophilicity [24]. It is thus expected that the *in situ* growth of LDHs nanosheets on this kind of carbon materials could largely improve the electrochemical performance of LDHs nanosheets.

Herein, we successfully synthesized porous carbon materials with hierarchical porous structure from abundant biomass wastes, *Dicranopteris dichotoma*. Subsequently, the NiCo-LDHs nanosheets were *in situ* synthesized and grown on the biomass waste-derived porous carbon (BC) to achieve hydrophilic NiCo-LDHs/BC materials with ultra-high charge storage capacity for high-performance HSCs (Fig. 1). To the best of our knowledge, no LDHs nanosheets anchored biomass waste-derived porous carbon are investigated as cathode materials for HSCs. This strategy exhibits the following advantages: (1) The *in situ* growth approach does not involve complicated manufacturing processes and any toxic organic solvents. (2) Ultrasonic treatment facilitates fast nucleation and enables controllable growth of LDHs nanosheets on the surface of BC with high specific surface area and hierarchical porous structure, which is beneficial to expose more active sites for charge storage. (3) The incorporation of BC materials improves the electronic conductivity and enhances the wettability of NiCo-LDHs/BC materials, which promote the rate capability and redox charge transfer ability. Furthermore, the effects of BC amount on the morphology and supercapacitive behavior of NiCo-LDHs/BC materials were studied. The aim of this work is to provide a new way to fabricate high-performance LDHs nanosheets based cathode materials for HSCs.

## 2. Experimental section

### 2.1. Synthesis of BC

All chemical materials were purchased and used without any treatment and purification. *Dicranopteris dichotoma* stems were washed with ultrapure water (18.25 MΩ cm) and dried at 120 °C for overnight. Then the dried stems were ground into powders using ball-milling and filtered through a 150 mesh stainless steel sieve. Typically, 1.0 g *Dicranopteris dichotoma* powders, 3.0 g potassium bicarbonate (KHCO$_3$, ACS, 99.7%–100.5%, Aladdin), and 1.0 g urea (AR, 99%, Macklin) were mixed in 20 mL ammonia solution (AR, GHTECH) for 24 h. Afterwards, the mixture was transferred to a 100 mL stainless steel autoclave for the hydrothermal reaction at 180 °C for 24 h. Subsequently, the mixture was activated in a tube furnace at 120 °C for 3 h, 450 °C for 3 h, and 750 °C for 3 h under a nitrogen atmosphere. After cooling down naturally, the resulting powders were ultrasonicated in 2 M hydrochloric acid (AR, Guangzhou Chemical Reagent Factory) for 1 h, and then washed with ultrapure water and ethanol (AR, ≥ 99.7%, GHTECH) repeatedly until the pH value of filtrate was 7. Finally, BC powders were obtained after vacuum drying at 60 °C for 12 h.

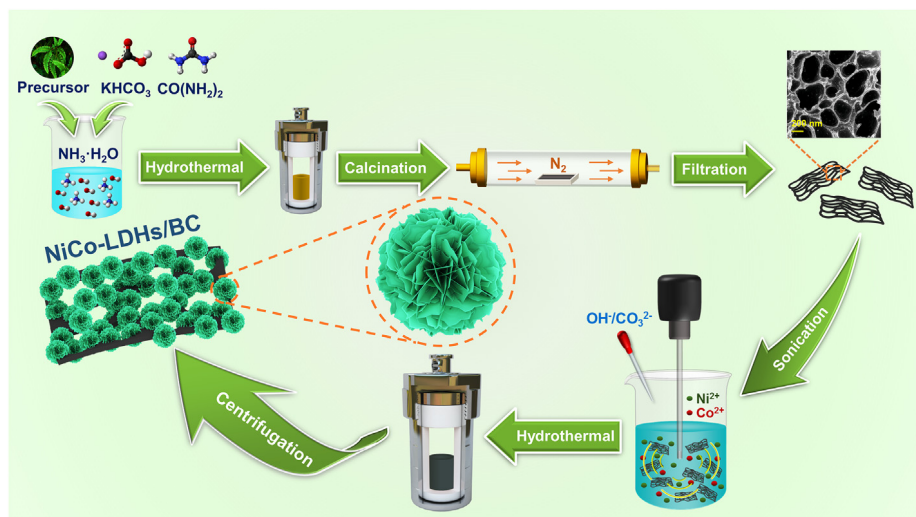

**Fig. 1.** Synthesis processes of NiCo-LDHs/BC materials.

## 2.2. Synthesis of NiCo-LDHs/BC

Prior to synthesis of NiCo-LDHs/BC materials, 1 M Ni$^{2+}$ and 1 M Co$^{2+}$ solutions were prepared by dissolving nickel chloride hexahydrate (NiCl$_2$·6H$_2$O, AR, 99%, Macklin) and cobalt chloride hexahydrate (CoCl$_2$·6H$_2$O, 98%, Alfa Asear) in ultrapure water, respectively. Firstly, BC powders were dispersed in 100 mL ultrapure water/ethanol (1:1) mixed solvent under ultrasonication for 30 min. Then, 4 mL of 1 M Ni$^{2+}$ solution and 4 mL of 1 M Co$^{2+}$ solution was added under ultrasonication for another 30 min. Afterwards, 0.8 M sodium hydroxide (NaOH, AR, 96%, Aladdin) and 0.2 M sodium carbonate (Na$_2$CO$_3$, GR, ≥ 99.9%, Macklin) aqueous mixed solution was dropped slowly until the pH value was 8.5. In the titration process, ultrasonication (40 kHz) and mechanical stirring (300 rpm) treatments were applied simultaneously for exfoliating LDHs. After ultrasonication/mechanical stirring for 1 h, the resulting solution was transferred and sealed for the hydrothermal treatment at 120 °C for 3 h. Finally, NiCo-LDHs/BC powders were obtained after centrifugation and vacuum drying at 80 °C for overnight. In this work, specific BC mass (0 mg, 25 mg, 50 mg, 75 mg) was controlled and the corresponding samples were denoted as Ni$_2$Co$_1$-LDHs/BC0, Ni$_2$Co$_1$-LDHs/BC25, Ni$_2$Co$_1$-LDHs/BC50, and Ni$_2$Co$_1$-LDHs/BC75. For comparison, NiCo-LDHs/BC with different molar ratios of nickel and cobalt (Ni$_1$Co$_1$-LDHs/BC50, Ni$_3$Co$_1$-LDHs/BC50) were also prepared following the same experimental procedures.

## 2.3. Synthesis of NiCo-LDHs thin films

NiCo-LDHs thin films were prepared by electrodeposition. The electrodeposition process was proceeded using an electrochemical workstation (Multi Autolab M204) with a three-electrode configuration. A nickel foam (MTI Corporation), a platinum filament (CHI115), and a Hg/HgO electrode (GaossUnion) were served as working electrode, counter electrode, and reference electrode, respectively. The electrolyte for electrodeposition was a mixed aqueous solution with the composition of 0.1 M Ni$^{2+}$ and 0.05 M Co$^{2+}$. Typically, the galvanostatic deposition was performed by applying the current density of 0.1 A cm$^{-2}$ for 2 min. The synthesized NiCo-LDHs thin films were rinsed with ultrapure water and vacuum dried at 60 °C for 12 h. The thin-film materials were denoted as Ni$_2$Co$_1$-LDHs-TF.

## 2.4. Characterizations

Material and electrochemical characterization methods of BC and NiCo-LDHs/BC materials are given in Supporting Information.

## 3. Results and discussion

### 3.1. Structural and morphological characterization

Fig. 2a displays the optical images of the NiCo-LDHs/BC materials. The color of as-prepared materials changes from initial deep green to black with the addition of BC. The crystal structures of BC and NiCo-LDHs/BC materials were determined by X-ray diffraction (XRD), which are shown in Fig. 2b and c, respectively. In Fig. 2b, a broad peak at around 24° corresponds to the (002) crystal plane, revealing the amorphous porous structure of BC materials [25]. In Fig. 2c, the diffraction peaks at around 34° and 61° can be ascribed to the (012) and (110) crystal planes of $Ni_2Co_1$-LDHs (PDF#33–0429), respectively [26]. The inapparent peaks related to (003) and (006) crystal planes indicate the effect of ultrasonication on the formation of LDHs nanosheets [16]. Meanwhile, the characteristic diffraction peaks of BC materials are not observed, demonstrating a homogeneous dispersion of porous BC in the resulting composites [27]. Accordingly, the XRD patterns manifest the successful preparation of hydrotalcite-like NiCo-LDHs nanosheets/BC composite materials.

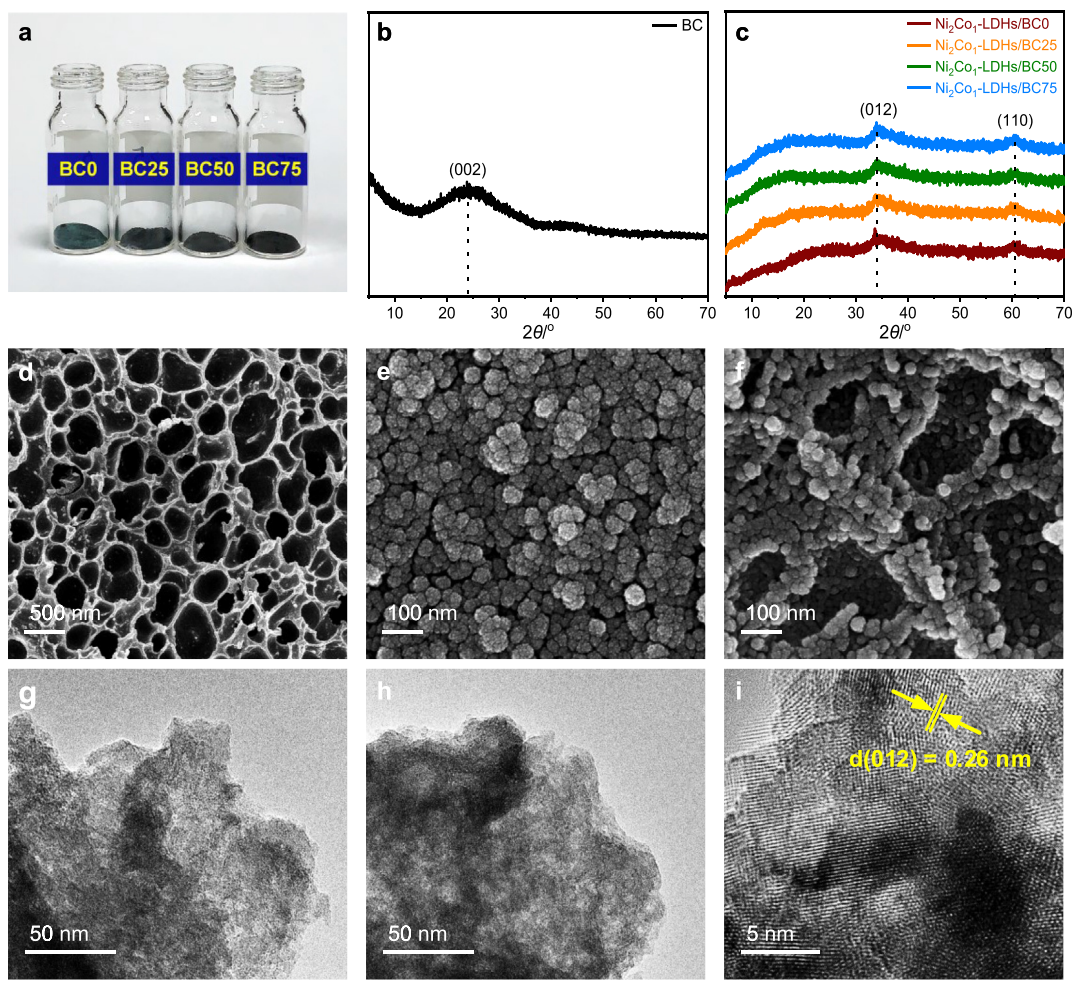

**Fig. 2.** (a) Optical images of Ni$_2$Co$_1$-LDHs/BC0, Ni$_2$Co$_1$-LDHs/BC25, Ni$_2$Co$_1$-LDHs/BC50, and Ni$_2$Co$_1$-LDHs/BC75 powders. XRD patterns of (b) BC, (c) Ni$_2$Co$_1$-LDHs/BC0, Ni$_2$Co$_1$-LDHs/BC25, Ni$_2$Co$_1$-LDHs/BC50, and Ni$_2$Co$_1$-LDHs/BC75. FE-SEM images of (d) BC, (e) Ni$_2$Co$_1$-LDHs/BC0, and (f) Ni$_2$Co$_1$-LDHs/BC50. TEM images of (g) Ni$_2$Co$_1$-LDHs/BC0 and (h) Ni$_2$Co$_1$-LDHs/BC50. (i) HRTEM image of Ni$_2$Co$_1$-LDHs/BC50.

Thermogravimetric analysis (TGA) measurements were also carried out to study the compositions of NiCo-LDHs/BC materials. As seen from TGA curves (Fig. S1), all NiCo-LDHs/BC samples exhibit a typical profile of LDHs with two distinct regions. The first region in the range of 30–180 °C is related to the elimination of absorbed water molecules [28]. The second region in the range of 180–340 °C is ascribed to the loss of interlayer water molecules and decomposition of interlayer carbonates [29]. Above 340 °C, $Ni_2Co_1$-LDHs/BC25, $Ni_2Co_1$-LDHs/BC50, and $Ni_2Co_1$-LDHs/BC75 show a rapid increase of weight loss, which corresponds to the combustion of carbon components [30]. The weight losses of these materials in the final stage are estimated as 6%, 7%, 10%, and 12%, respectively.

Field emission scanning electron microscopy (FE-SEM) was performed to study the morphologies of BC and NiCo-LDHs/BC materials. As shown in Fig. 2d, BC materials present a distinct 3D porous framework with randomly macropores in the range of hundreds of nanometers. As a reference material, the pristine NiCo-LDHs display a uniform microspherical morphology (Fig. 2e). Noticeably, the average diameter of NiCo-LDHs microspheres is only 40–50 nm with the aid of ultrasonication. From Fig. 2f, and Fig. S2, NiCo-LDHs microspheres are successfully anchored on the porous skeleton of BC by applying an *in situ* growth approach. With the increasing BC contents, carbon frameworks are covered by more NiCo-LDHs microspheres. When BC content is 50 mg, a uniform distribution of NiCo-LDHs microspheres is distinctly seen on the surface of BC. The well-defined macropores are beneficial for the rapid penetration of electrolyte ions. Meanwhile, the energy dispersive X-ray spectroscopy (EDX) mappings (Fig. S3) also prove the homogeneous distribution of NiCo-LDHs microspheres throughout the $Ni_2Co_1$-LDHs/BC50 materials. However, for $Ni_2Co_1$-LDHs/BC75 materials, it is clearly observed that the macropores are partially blocked by NiCo-LDHs microspheres, giving rise to the reduced surface area.

Furthermore, the morphology of NiCo-LDHs microspheres in as-synthesized materials was studied using transmission electron microscopy (TEM), atomic force microscopy (AFM) technologies. The transparent edges in TEM images (Fig. 2g and h) confirm the formation of ultrathin nanosheets in NiCo-LDHs/BC materials. The high-resolution transmission electron microscopy (HR-TEM) image (Fig. 2i) displays the lattice fringes with a *d*-spacing of 0.26 nm, which corresponds to the (012) crystal planes of NiCo-LDHs in XRD results. Interestingly, according to Fig. 3, the thickness values of NiCo-LDHs nanosheets in $Ni_2Co_1$-LDHs/BC0 and $Ni_2Co_1$-LDHs/BC50 samples are calibrated as 5.20 nm and 6.20 nm, respectively, denoting that the ultrathin characters of LDHs nanosheets are not affected with the addition of porous carbon.

The porosity parameters including specific surface area and pore volume of as-prepared materials were characterized by the nitrogen adsorption/desorption isotherms. For BC, the nitrogen adsorption/desorption isotherm (Fig. 4a) displays a typical type I curve, indicating the microporous dominated structure [31]. The corresponding pore size distribution in Fig. 4b results in the micropores have a maximum peak at 0.80 nm. For NiCo-LDHs/BC materials, Fig. 4c presents the nitrogen adsorption/desorption isotherms with the combined type I and type IV characteristics [32]. All NiCo-LDHs/BC materials are composed of micropores and mesopores, which are confirmed by the pore size distribution in Fig. 4d. Combining the FE-SEM images, it is obvious that NiCo-LDHs/BC materials possess the hierarchical pore structure with abundant micropores, mesopores, and macropores. The calculated porosity parameters are illustrated in Table 1. The remarkable high Brunauer-Emmett-Teller (BET) specific surface area of KHCO$_3$ activated BC is achieved as 2324.1 m$^2$ g$^{-1}$, which is comparable to or superior to the recent publications [33,34]. As a result, the incorporation of BC into NiCo-LDHs improves the BET specific surface area and total pore volume significantly. The maximum BET specific surface area and total pore volume of NiCo-LDHs/BC materials can reach 168.0 m$^2$ g$^{-1}$ and 0.149 cm$^3$ g$^{-1}$, respectively. Moreover, the largest microporous specific surface area and highest mesoporous/macroporous specific surface area of Ni$_2$Co$_1$-LDHs/BC50 provide abundant active sites for charge storage and plenty of channels for ion transport, which is essential for excellent capacitive performance [35].

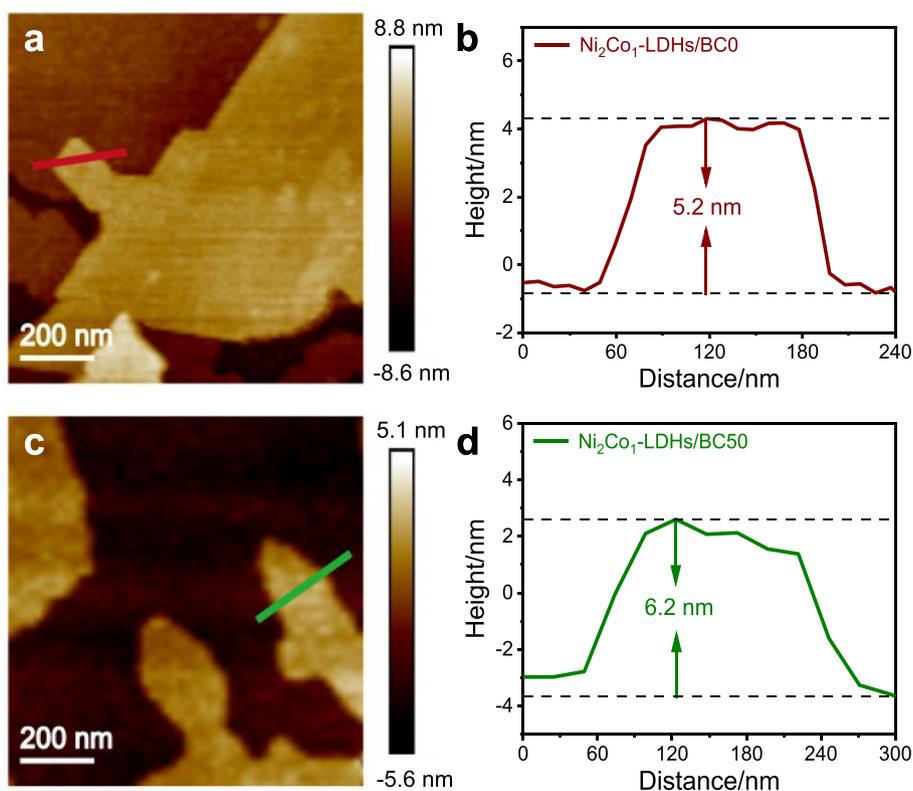

**Fig. 3.** AFM images and corresponding height profiles of (a, b) $Ni_2Co_1$-LDHs/BC0 and (c, d) $Ni_2Co_1$-LDHs/BC50.

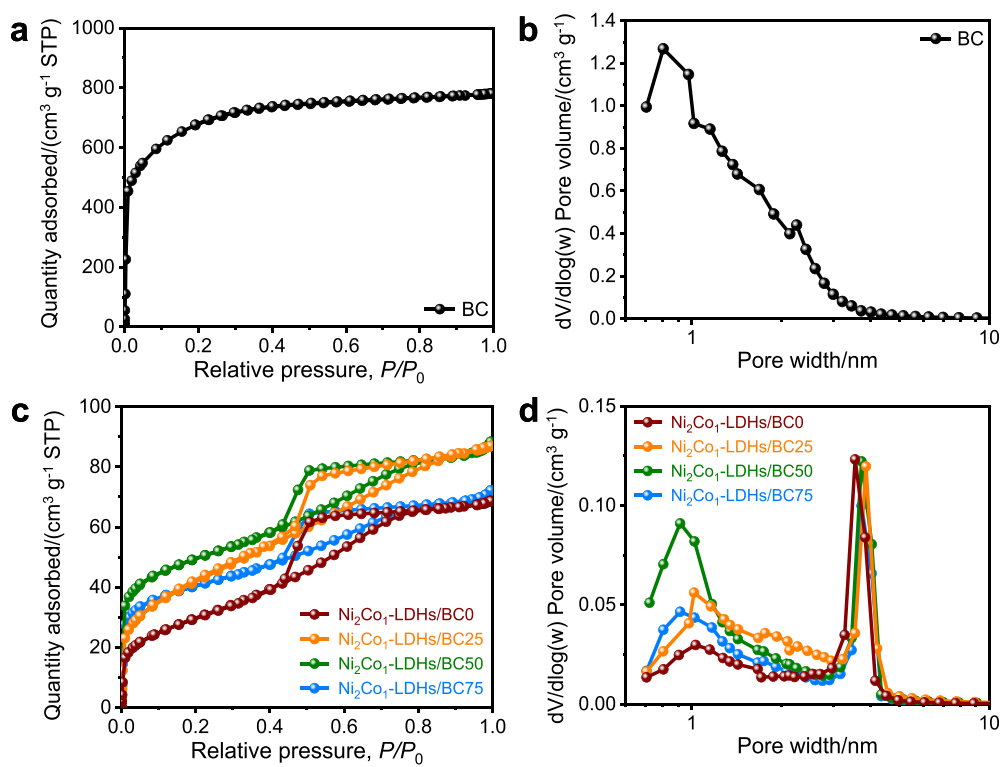

**Fig. 4.** (a) Nitrogen adsorption/desorption isotherm and (b) pore size distribution of BC. (c) Nitrogen adsorption/desorption isotherms and (d) pore size distribution of Ni$_2$Co$_1$-LDHs/BC0, Ni$_2$Co$_1$-LDHs/BC25, Ni$_2$Co$_1$-LDHs/BC50, and Ni$_2$Co$_1$-LDHs/BC75.

**Table 1**
Porosity parameters of BC, Ni$_2$Co$_1$-LDHs/BC0, Ni$_2$Co$_1$-LDHs/BC25, Ni$_2$Co$_1$-LDHs/BC50, and Ni$_2$Co$_1$-LDHs/BC75.

| Samples | $S_{BET}/(m^2\ g^{-1})$ | $S_{micro}/(m^2\ g^{-1})$ | $S_{meso\ +\ macro}/(m^2\ g^{-1})$ | $V_{total}/(cm^3\ g^{-1})$ | $V_{micro}/(cm^3\ g^{-1})$ | $V_{meso\ +\ macro}/(cm^3\ g^{-1})$ |
|---|---|---|---|---|---|---|
| BC | 2324.1 | 1800.0 | 524.1 | 1.407 | 0.499 | 0.908 |
| Ni$_2$Co$_1$-LDHs/BC0 | 106.6 | 22.6 | 84.0 | 0.089 | 0.002 | 0.087 |
| Ni$_2$Co$_1$-LDHs/BC25 | 149.3 | 45.3 | 104.0 | 0.132 | 0.024 | 0.108 |
| Ni$_2$Co$_1$-LDHs/BC50 | 168.0 | 58.9 | 109.1 | 0.149 | 0.029 | 0.120 |
| Ni$_2$Co$_1$-LDHs/BC75 | 137.3 | 48.1 | 89.2 | 0.121 | 0.023 | 0.098 |

### 3.2. Electrochemical characterization

The capacitive performance of as-prepared BC and NiCo-LDHs/BC electrodes were evaluated using CV, GCD, EIS with a three-electrode system in 6 M potassium hydroxide (KOH) electrolyte. Fig. 5a shows CV curves of BC electrode at different scan rates of 5–100 mV s$^{-1}$. The CV curves with a quasi-rectangular shape reveal a classical electrical double layer capacitive behavior [36]. The GCD curves of BC electrode in Fig. S4a exhibit the symmetrical triangular shape with negligible IR drop, illuminating excellent reversibility. According to GCD curves, BC electrode displays the maximum specific capacitance of 310 F g$^{-1}$ at 0.5 A g$^{-1}$ and high capacitance retention of 76% from 0.5 A g$^{-1}$ to 10 A g$^{-1}$, which are illustrated in Fig. S4b.

Fig. 5b presents the CV curves at a scan rate of 50 mV s$^{-1}$ of NiCo-LDHs/BC electrodes. The CV curves have a pair of prominent redox peaks related to the transformation of Ni(OH)$_2$/NiOOH and Co(OH)$_2$/CoOOH/CoO$_2$ [37]. Specifically, Ni$_2$Co$_1$-LDHs/BC50 electrode with the suitable hierarchically porous structure possesses the greatest specific current. Meanwhile, with the increasing scan rates, the CV curves of Ni$_2$Co$_1$-LDHs/BC50 electrode in Fig. 5c remain similar shapes with pronounced shifts of redox peaks because of internal resistance and the nature of Faradaic redox reactions [38]. Based on these CV curves, the electrochemical charge storage kinetics analysis is proceeded using the equation between peak current ($i$) and scan rate ($\nu$): $i = a\nu^b$ [39]. Typically, a $b$-value of 0.5 indicates a diffusion-controlled Faradaic process and a $b$-value of 1.0 suggests a capacitive process. By fitting in Fig. S5, the $b$-value is determined as 0.71, demonstrating that the Ni$_2$Co$_1$-o$_1$-LDHs/BC50 electrode is governed by both mechanisms. Specifically, two mechanisms are separated by the formula: $i = k_1\nu + k_2\nu^{0.5}$ [39]. From Fig. 5d, the Ni$_2$Co$_1$-LDHs/BC50 electrode shows negligible the capacitive contribution (16.7%) at 5 mV s$^{-1}$. With the increasing scan rate, the capacitive contribution increases rapidly to 86% at 50 mV s$^{-1}$,

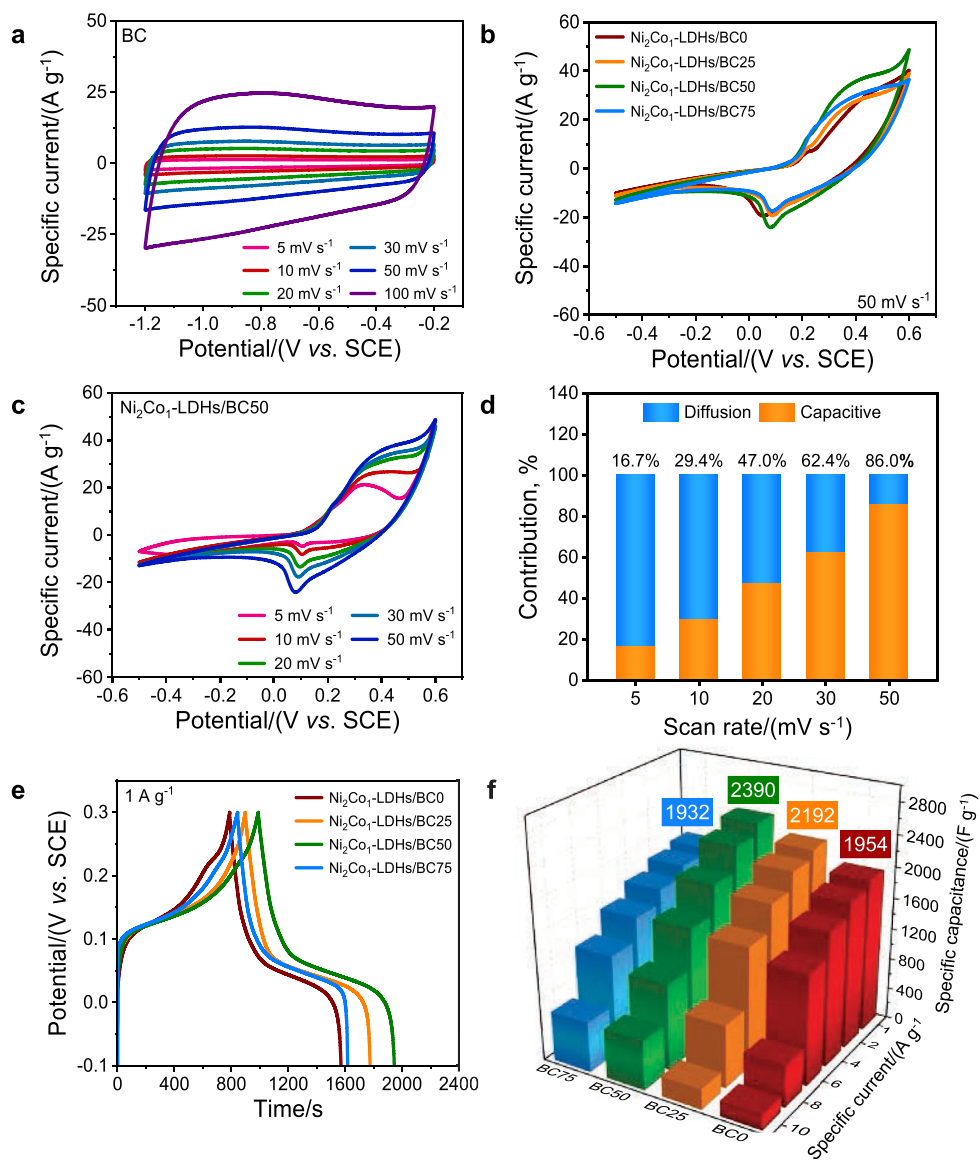

**Fig. 5.** (a) CV curves of the BC electrode at different scan rates. (b) CV curves of the Ni₂Co₁-LDHs/BC0, Ni₂Co₁-LDHs/BC25, Ni₂Co₁-LDHs/BC50, and Ni₂Co₁-LDHs/BC75 electrode at a scan rate of 50 mV s⁻¹. (c) CV curves of the Ni₂Co₁-LDHs/BC50 electrode at different scan rates. (d) Contribution ratio of the capacitive and diffusion-controlled mechanism at different scan rates for the Ni₂Co₁-LDHs/BC50 electrode. (e) GCD curves of the Ni₂Co₁-LDHs/BC0, Ni₂Co₁-LDHs/BC25, Ni₂Co₁-LDHs/BC50, and Ni₂Co₁-LDHs/BC75 electrode at a specific current of 1 A g⁻¹. (f) Specific capacitance values of NiCo-LDHs/BC electrode at different specific currents.

reflecting that the Ni₂Co₁-LDHs/BC50 electrode relies on capacitive mechanism at high scan rates.

Fig. 5e illustrates GCD curves of NiCo-LDHs/BC electrodes at a specific current of 1 A g⁻¹. Ni₂Co₁-LDHs/BC50 electrode gives the longest discharge time of 956 s, whereas the discharge time of Ni₂Co₁-LDHs/BC0 electrode, Ni₂Co₁-LDHs/BC25 electrode, and Ni₂Co₁-LDHs/BC75 electrode is 782 s, 877 s, and 773 s, respectively. GCD curves of all Ni₂Co₁-LDHs/BC electrodes in Fig. S6 maintain symmetrical at different specific currents, revealing great columbic efficiency. Subsequently, the specific capacitance values of NiCo-LDHs/BC electrodes are calculated based on these GCD curves at various specific currents. As seen in Fig. 5f, the maximum specific capacitance of NiCo-LDHs/BC electrodes reaches 2390 F g⁻¹ (956 C g⁻¹) at 1 A g⁻¹, which is 1.2 times that of pristine NiCo-LDHs electrode. Moreover, from 1 A g⁻¹ to 10 A g⁻¹, the specific capacitance value of NiCo-LDHs/BC0 electrode and NiCo-LDHs/BC75 electrode retain 10% and 34%, respectively, demonstrating the improvement of rate capability with the introduction of BC.

To further understand the effect of carbon components, EIS analyses of NiCo-LDHs/BC electrodes were carried out in the frequency range from 0.01 Hz to 100 kHz as depicted in Fig. S7. After fitting the EIS data with an equivalent circuit, the equivalent series resistance $R_{ESR}$ and the charge transfer resistance $R_{ct}$ are displayed in Table S1. Compared with pristine NiCo-LDHs electrode, the decreasing $R_{ESR}$ and $R_{ct}$ values of NiCo-LDHs/BC electrodes demonstrate that the incorporation of BC components could accelerate electron transfer and facilitate ion transport dramatically, owing to the intrinsic high electronic conductivity of BC and the reduced agglomeration of NiCo-LDHs nanosheets [22]. Afterwards, the capacitive performance of Ni₁Co₁-LDHs/BC50 and Ni₃Co₁-LDHs/BC50 electrode were also investigated. In Fig. S8, with the increasing Ni/Co ratio, the charge/discharge platform elevates owing to the inherent electrochemical properties of Ni and Co [40]. The specific capacitance of Ni₁Co₁-LDHs/BC50 and Ni₃Co₁-LDHs/BC50 is 1766 F g⁻¹ (706 C g⁻¹) and 1692 F g⁻¹ (677 C g⁻¹) at 1 A g⁻¹, respectively, indicating that the best Ni/Co ratio of NiCo-LDHs/BC materials for realizing high capacitance is 2:1.

The high electrochemical performance of NiCo-LDHs/BC materials in this work can be ascribed to the synergistic effect of the proper arrangement of ultrathin LDHs nanosheets, large specific surface area, and high conductivity with the addition of BC. Except for above considerable reasons, improvement of the wettability is also proposed to disclose the reasons for the increased specific capacitance. A controlled experiment was designed to compare the specific capacitance of Ni₂Co₁-

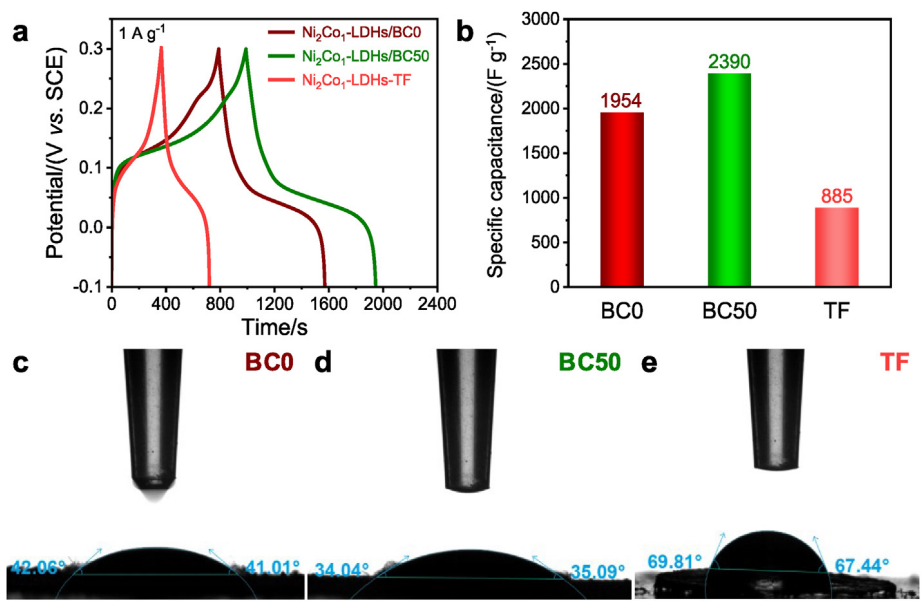

**Fig. 6.** (a) GCD curves and (b) corresponding specific capacitance of the Ni₂Co₁-LDHs/BC0, Ni₂Co₁-LDHs/BC50, and Ni₂Co₁-LDHs-TF electrode at 1 A g⁻¹. Contact angle measurements of (c) Ni₂Co₁-LDHs/BC0, (d) Ni₂Co₁-LDHs/BC50, and (e) Ni₂Co₁-LDHs-TF with 6 M KOH electrolyte.

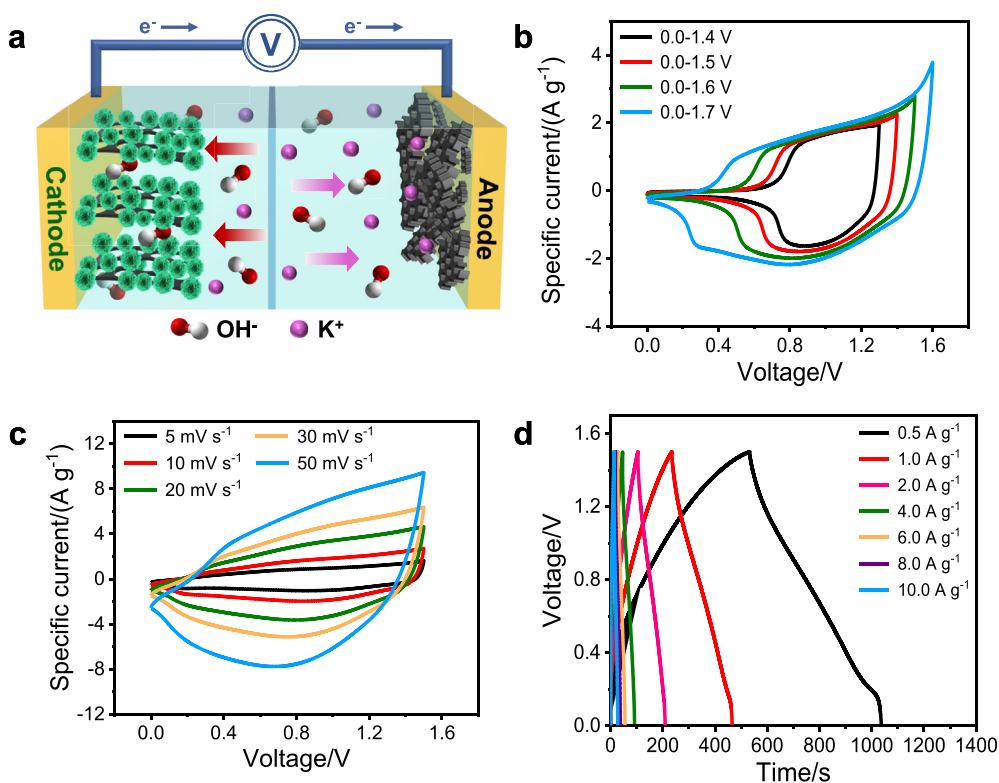

**Fig. 7.** (a) Schematic illustration of the as-fabricated HSC device. (b) CV curves of the HSC performed at different voltage windows at 10 mV s⁻¹. (c) CV curves of the HSC at different scan rates from 5 to 50 mV s⁻¹. (d) GCD curves of the HSC at different specific currents from 0.5 to 10.0 A g⁻¹.

LDHs/BC0, Ni₂Co₁-LDHs/BC50, and Ni₂Co₁-LDHs-TF. As shown in Fig. 6, Ni₂Co₁-LDHs/BC50 with the highest specific capacitance of 2390 F g⁻¹ exhibits the contact angles of 34°–35° with 6 M KOH electrolyte, revealing a much more electrolyte-hydrophilic surface. In contrast, the contact angle of Ni₂Co₁-LDHs/BC0 and Ni₂Co₁-LDHs-TF decreases to 41°–42° and 67°–70°, which corresponds to the specific capacitance of 1954 F g⁻¹ (782 C g⁻¹) and 885 F g⁻¹ (354 C g⁻¹), respectively. The improved surface wettability accelerates the exposure rates of the interaction between the hydroxide ions of the electrolyte and Ni²⁺/Co²⁺, which ensures rapid electrolyte ion penetration and enhanced utilization of active sites of LDHs nanosheets [41]. This trend is also in good agreement with the EIS results. Thus the capacitive performance of Ni₂Co₁-LDHs/BC50 is superior to that of NiCo-LDHs in relative publications (Table S2).

To evaluate the practical applications of the Ni₂Co₁-LDHs/BC50 electrode materials, a HSC was fabricated using Ni₂Co₁-LDHs/BC50 as the positive electrode material and commercial activated carbon YP-80F as negative electrode material (Fig. 7a). Fig. S9a presents three-electrode CV curves of the Ni₂Co₁-LDHs/BC50 electrode and the YP-80F electrode. The corresponding operating potential is determined as -1.0–0.0 V and -0.3–0.5 V, respectively. Thus the voltage window of the HSC is optimized as 1.5 V, which is demonstrated in Fig. 7b. After 1.5 V, the irreversible decomposition of water splitting is pronounced, resulting in an asymmetric CV curve [42]. The electrochemical performance of the HSC was investigated by CV curves at various scan rates (5–50 mV s⁻¹) and GCD curves at various specific currents (0.5–10.0 A g⁻¹). As illustrated in Fig. 7c, all CV curves exhibit symmetrical broad redox peaks, which originate from the electrochemical process of positive electrode material. Accordingly, the GCD curves in Fig. 7d possess the distinguished electrochemical reversibility. Based on GCD curves, Fig. S9b illustrates that the maximum specific capacitance of the HSC is calculated as 168 F g⁻¹ at 0.5 A g⁻¹. When the specific current increases to 10 A g⁻¹, the specific capacitance of the HSC is still maintained at 92 F g⁻¹, showing an acceptable capacitance retention of 55%.

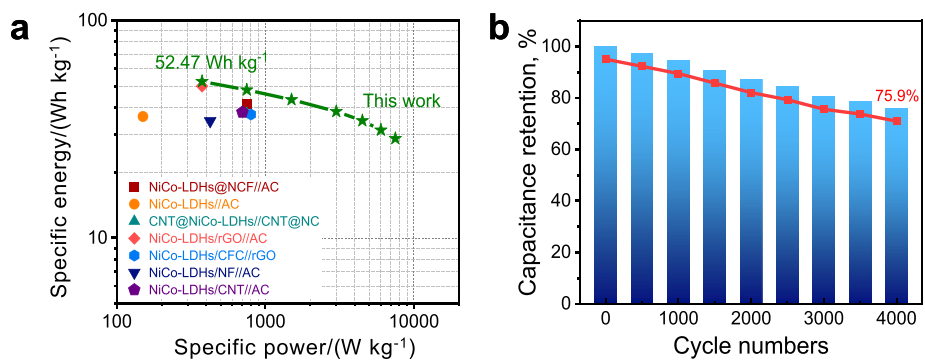

**Fig. 8.** (a) Ragone plot related to the specific energy and specific power of the HSC. (b) Cycle stability of the HSC at 10 A g⁻¹.

The Ragone plot is a functional tool for comparing the electrochemical performance of various energy storage devices. As depicted in Fig. 8a, the Ni₂Co₁-LDHs-BC50//YP80F HSC achieved a high specific energy of 52.47 Wh kg⁻¹ at a specific power of 375 W kg⁻¹ and a specific energy of 28.66 Wh kg⁻¹ at a high specific power of 7500 W kg⁻¹. The maximum specific energy value of the HSC in this work is comparable or superior to those of the NiCo-LDHs based HSCs in recent literatures, such as NiCo-LDHs@NCF//AC HSC (41.5 Wh kg⁻¹ at 750 W kg⁻¹) [43], NiCo-LDHs//AC HSC (36.2 Wh kg⁻¹ at 150 W kg⁻¹) [37], CNT@NiCo-LDHs//CNT@NC HSC (37.4 Wh kg⁻¹ at 750 W kg⁻¹) [44], NiCo-LDHs/rGO//AC HSC (49.9 Wh kg⁻¹ at 375 W kg⁻¹) [45], NiCo-LDHs/CFC//rGO HSC (37.0 Wh kg⁻¹ at 800 W kg⁻¹) [40], NiCo-LDHs/NF//AC HSC (34.5 Wh kg⁻¹ at 425 W kg⁻¹) [46], NiCo-LDHs/CNT//AC HSC (37.9 Wh kg⁻¹ at 701 W kg⁻¹) [47]. Besides, the HSC exhibits excellent cycling performance with the capacitance retention of 75.9% after 4000 cycles (Fig. 8b) due to the enhanced wettability of cathode materials [48]. The morphology of the Ni₂Co₁-LDHs/BC50 electrode before and after cycling (Fig. S10) is well maintained, indicating its good cycling stability.

## 4. Conclusions

In summary, we successfully synthesized hydrophilic NiCo-LDHs nanosheets on biomass waste-derived porous carbon (BC) framework with a facile *in situ* growth method. The *in situ* growth could greatly prevent the aggregation of LDHs nanosheets, and further improve their specific surface area, electronic conductivity, and wettability. The rational arrangement of NiCo-LDHs nanosheets on the surface of BC gives rise to a remarkable specific capacitance of 2390 F g⁻¹ (956 C g⁻¹) at 1 A g⁻¹. Besides, a HSC using Ni₂Co₁-LDHs/BC50 as the positive electrode and YP-80F as the negative electrode manifests a high specific energy of 52.47 Wh kg⁻¹ at 375 W kg⁻¹, which is superior to most candidates previously reported. In addition, the HSC shows excellent cycling stability, remaining 75.9% of original specific capacitance after 4000 cycles at 10 A g⁻¹. Finally, this work not only provides an economical synthesis method of LDHs nanosheets based electrode materials but also accelerates the prosperous development of energy storage devices.

## Declaration of competing interests

The authors declare that they have no known competing financial interests or personal relationships that could have appeared to influence the work reported in this paper.

## Acknowledgments

The authors acknowledged the financial supports from Key-Area Research and Development Program of Guangdong Province (2019B110209003), National Natural Science Foundation of China (21776324, 22078374), Guangdong Basic and Applied Basic Research Foundation (2019B1515120058, 2020A1515011149), National Key R&D Program of China (2018YFD0800703, 2020YFC1807600), National Ten Thousand Talent Plan, the Fundamental Research Funds for the Central Universities (19lgzd25), and Hundred Talent Plan (201602) from Sun Yat-sen University.

## Appendix A. Supplementary data

Supplementary data to this article can be found online at https://doi.org/10.1016/j.gce.2021.09.001.